\documentclass[12pt]{article}
\usepackage{epsfig}
\usepackage{amssymb}
\usepackage{amsmath,amsfonts,amssymb,graphicx}
\newcommand {\be}{\begin{equation}}
\newcommand {\ee}{\end {equation}}
\newcommand{\beq}{\begin{eqnarray}}
\newcommand{\eeq}{\end{eqnarray}}

\begin{document}
\title{Neutrino Oscillations With Two Sterile Neutrinos}
\author{Leonard S. Kisslinger\\
Department of Physics, Carnegie Mellon University, Pittsburgh PA 15213}
\maketitle
\date{}
\noindent
PACS Indices:11.30.Er,14.60.Lm,13.15.+g
\vspace{3mm}


\begin{abstract}
 This work estimates the probability  of $\mu$ to $e$ neutrino oscillation 
with two sterile neutrinos using a 5x5 U-matrix, an extension of the previous
estimate with one sterile neutrino and a 4x4 U-matrix. The sterile 
neutrino-active neutrino mass differences and the mixing angles of the two 
sterile neutrinos with the three active neutrinos are taken from recent 
publications, and the oscillation probability for one sterile neutrino
is compared to the previous estimate.
\end{abstract}

\section{Introduction}

Reviews of experimental data on neutrino oscillations\cite{kms11,kmms13,
ggllz15} find that there probably are two sterile neutrinos. 
Refs. \cite{kms11,kmms13} by Kopp $et.\;al.$ discuss a variety of experiments 
on neutrino oscillations, with appearance and disappearance, while the present 
work treats $\nu_\mu$ to $\nu_e$ appearance. A recent analysis
of neutrino oscillation experiments with one and two sterile 
neutrinos\cite{collin16} estimate the sterile neutrino masses and mixing 
angles used in the present work.

In the present work we use a U-matrix approach, introduced for active
neutrinos  with a 3x3 U-matrix\cite{as97}, and extended to a 4x4 U-matrix
with one sterile neutrino in a recent study of $\mathcal{P}(\nu_\mu
\rightarrow \nu_e)$, the transition probability for a muon
neutrino to oscillate to an electron neutrino\cite{lsk14,lsk15}. We introduce a
5x5 U-matrix for three active and two sterile neutrinos, using the parameters
found in Refs.\cite{kms11,kmms13} and Ref.\cite{collin16}
 
\section{ 5x5 U-Matrix}

Active neutrinos with flavors $\nu_e,\nu_\mu,\nu_\tau$ and two sterile 
neutrinos, $\nu_{s_1},\nu_{s_2}$ are related to neutrinos with 
definite mass by
\beq
\label{f-mrelation}
      \nu_f &=& U\nu_m \; ,
\eeq
where $U$ is a 5x5 matrix and $\nu_f,\nu_m$ are 5x1 column vectors.
We use the notation  $s_{ij}, c_{ij}=sin\theta_{ij},cos\theta_{ij}$, with 
$\theta_{12}, \theta_{23}, \theta_{13}$ the mixing angles for active neutrinos; 
and $s_\alpha=sin(\alpha), c_\alpha=cos(\alpha), s_\beta=sin(\beta),c_\beta=
cos(\beta)$, where $\alpha=\theta_{i4},\beta=\theta_{i5}$ are sterile-active 
neutrino mixing angles, with i=1,2,3, and $\delta_{CP}$=0.
\beq
\label{Uform}
    U &=& O^{23}O^{13} O^{12} O^{14} O^{24} O^{34} O^{15} O^{25} O^{35} 
O^{45} \; , 
\eeq
where ($O^{45}$, giving sterile-sterile neutrino mixing, is not shown)
\vspace{3mm}

$O^{23}$=
 $\left( \begin{array}{ccclcr} 1 & 0 & 0 & 0 & 0 \\ 0 & c_{23} & s_{23} & 0 
& 0\\
0 & -s_{23} & c_{23} & 0 & 0\\ 0 & 0 & 0 & 1 & 0\\ 0 & 0 & 0 & 0 & 1 \end{array} 
\right)$
\hspace{3mm}$O^{13}$=
$\left( \begin{array}{ccclcr} c_{13} & 0 & s_{13} & 0 & 0\\ 0 & 1 & 0 & 0 
& 0 \\-s_{13} & 0  & c_{13} & 0 & 0 \\ 0 & 0 & 0 & 1 & 0\\ 0 & 0 & 
0 & 0 & 1  \end{array} \right)$
\vspace{3mm}

$O^{12}$=
 $\left( \begin{array}{ccclcr} c_{12} & s_{12} & 0 & 0 & 0\\ -s_{12} & 
c_{12} & 0 & 0 & 0 \\ 0 & 0  & 1 & 0 & 0 \\ 0 & 0 & 0 & 1 & 0\\
0 & 0 & 0 & 0 & 1 \end{array} \right)$
\hspace{3mm}$O^{14}$=
 $\left( \begin{array}{ccclcr} c_\alpha & 0 & 0 & s_\alpha & 0\\ 
 0 & 1  & 0 & 0 & 0\\  0 & 0 & 1 & 0 & 0\\
 -s_\alpha & 0 & 0 & c_\alpha & 0\\
0& 0 & 0 & 0 & 1 \end{array} \right)$
\vspace{3mm}

$O^{24}$=
$ \left( \begin{array}{ccclcr} 1 & 0 & 0 & 0 & 0 \\ 0 & c_\alpha & 0 & 
s_\alpha & 0\\ 0 & 0 & 1 & 0 & 0\\ 0 & -s_\alpha & 0 & c_\alpha & 0 \\
 0 & 0  & 0 & 1 & 0\\
0& 0 & 0 & 0 & 1 \end{array} \right)$
\hspace{3mm}$O^{34}$=
$\left( \begin{array}{ccclcr} 1 & 0 & 0 & 0 & 0 \\ 0 & 1 & 0 & 0 & 0 \\
 0 & 0  & c_\alpha & s_\alpha & 0 \\ 0 & 0 & -s_\alpha & c_\alpha & 0\\
0& 0 & 0 & 0 & 1 \end{array} \right)$

$O^{15}$=
$\left( \begin{array}{ccclcr} c_\beta & 0 & 0 & 0 & s_\beta\\
 0 & 1 & 0 & 0 & 0\\ 0 & 0 & 1 & 0 & 0\\ 0 & 0 & 0 & 1 & 0\\ 
- s_\beta & 0 & 0 & 0 & c_\beta \end{array} 
\right)$
\hspace{3mm}$O^{25}$=
 $\left( \begin{array}{ccclcr}  1 & 0 & 0 & 0 & 0\\ 0 & c_\beta & 0 & 0 &
 s_\beta\\ 0 & 0 & 1 & 0 & 0\\ 0 & 0 & 0 & 1 & 0 \\ 
 0 &-s_\beta  & 0 & 0 & c_\beta \end{array}
\right)$
\vspace{3mm}

\hspace{3cm}$O^{35}$=
$\left( \begin{array}{ccclcr} 1 & 0 & 0 & 0 & 0\\ 0 & 1 & 0 & 0 & 0\\
 0 & 0 & c_\beta & 0 & s_\beta\\ 0 & 0 & 0 & 1 & 0\\ 0& 0 
 & -s_\beta  & 0 & c_\beta\end{array} \right)$

\vspace{5mm}

  $ \mathcal{P}(\nu_\mu \rightarrow\nu_e)$ 
is obtained from the 5x5 U matrix and the neutrino mass differences
$\delta m_{ij}^2=m_i^2-m_j^2$ for a neutrino beam with energy $E$ and baseline
$L$ by
\beq
\label{Pue-1}
 \mathcal{P}(\nu_\mu \rightarrow\nu_e) &=& Re[\sum_{i=1}^{5}\sum_{j=1}^{5}
U_{1i}U^*_{1j}U^*_{2i}U_{2j} e^{-i(\delta m_{ij}^2/E)L}] \; ,
\eeq
an extension of the 4x4\cite{lsk14,lsk15} theory with one serile neutrino, 
which used the  3x3 formalism of Ref\cite{as97}, to a 5x5 matrix 
formalism.
From Eq(\ref{Uform}), multiplying the nine 5x5 $O$ matrices, we obtain the 
matrix U. With $\delta_{CP}$=0, $U^*_{ij}=U_{ij}$, so we only need $U_{1j},U_{2j}$.
The active neutrino mixing parameters\cite{hjk11} are $c23=s23=.7071,c13=.989,
s13=.15,c12=.83,s12=.56$.
\beq
\label{U1j}
  U_{11}&=&.821 ca{\rm \;}cb  \nonumber \\
  U_{12} &=& (.554 ca - .821 sa^2) cb - .821 ca{\rm \;}sb^2  \nonumber \\
 U_{13}&=&(.15 ca-.554 sa^2-.821ca{\rm \;}sa^2)cb-(.554 ca - .821 sa^2)sb^2
\nonumber \\
      && +.821 ca{\rm \;}cb{\rm \;}sb^2\nonumber \\
 U_{14} &=&cb(.15sa +.554 ca{\rm \;}sa + .821 ca^2{\rm \;}sa)-.821ca{\rm \;}
cb^2{\rm \;}sb^2
\nonumber \\
   && -(.554 ca-.821 sa^2)cb{\rm \;}sb^2-(.15 ca-.554 sa^2-.821 ca sa^2)sb^2
\nonumber \\
U_{15} &=&.821ca{\rm \;}sb{\rm \;}cb^3+(.15sa+.554ca{\rm \;}sa+
.821 ca^2{\rm \;}sa)sb \nonumber \\
   && +(.554 ca-.821 sa^2)cb^2{\rm \;}sb+(.15ca-.554sa^2-.821ca{\rm \;}sa^2)
cb{\rm \;}sb \nonumber \\ 
 U_{21}&=&-.484ca{\rm \;}cb \\
  U_{22}&=&(.527ca+.484 sa^2)cb+.484ca{\rm \;}sb^2) \nonumber \\
  U_{23}&=& (.699ca-.527sa^2+.484ca{\rm \;}sa^2)cb-(.527ca+.484sa^2)sb^2 
+.484ca{\rm \;}cb{\rm \;}sb^2\nonumber \\
  U_{24}&=& cb(.699 sa+.527ca{\rm \;}sa-.484ca^2{\rm \;}sa)+
.484ca{\rm \;}cb^2{\rm \;}sb^2 \nonumber \\
   && -(.527ca +.484sa^2)cb{\rm \;}sb^2 -(.699ca-.527sa^2+.484ca{\rm \;}sa^2) 
sb^2 \nonumber \\
  U_{25}&=& -.484 ca{\rm \;}sb{\rm \;}cb^3 +(.699sa+.527ca{\rm \;}sa
-.484ca^2{\rm \;}sa)sb\nonumber \\
   && +(.527ca+.484sa^2)cb^2{\rm \;}sb +(.699ca -.527sa^2+.484 ca{\rm \;}sa^2) 
cb{\rm \;}sb
\nonumber
\eeq

  The active neutrino mass differences are $\delta m_{12}^2=m_2^2-m_1^2=
7.6 \times 10^{-5}(eV)^2$, $\delta m_{13}^2=m_3^2-m_1^2\simeq \delta m_{23}^2 = 
2.4\times 10^{-3} (eV)^2$. From Ref\cite{collin16} the  first sterile-active 
mass difference =  $\delta m_{4i}^2=m_4^2-m_i^2 \simeq$ 1.75 (eV)$^2$ ,
with i=1,2,3 for active neutrinos; and $s_\alpha^2 \simeq 2.6x10^{-2}$, or
$\alpha \simeq 9.2^o$. Because of the difficulty in the analyis we assume
that  $\delta m_{4i}^2= \delta m_{5i}^2$, $\delta m_{54}^2=0$ and $\alpha=\beta$. 
Note that the sterile-active mixing angle used in Refs\cite{lsk14,lsk15} was 
also $9.2^o$ .
\newpage

\section{$\mathcal{P}(\nu_\mu \rightarrow \nu_e)$ For Two Sterile
Neutrinos}

With the mass differences $\delta m_{12}^2$, $\delta m_{13}^2$, $\delta m_{23}^2$,
 $\delta m_{4i}^2$, $\delta m_{5i}^2$, $\delta m_{54}^2$ given above, we define 
$\delta=\delta m_{12}^2/2E$ , $\Delta=\delta m_{13}^2/2E$, $\gamma= 
\delta m_{4i}^2/2E$, $\lambda= \delta m_{5i}^2/2E$, $\kappa=\delta m_{54}^2/2E$.

\beq
\label{Pue-2}
\mathcal{P}(\nu_\mu \rightarrow \nu_{e}) &=&Re[U_{11}U_{21}( U_{11}U_{21}+
 U_{12}U_{22} e^{-i\delta L}+ U_{13}U_{23} e^{-i\Delta L}+ \nonumber \\
  && U_{14}U_{24} e^{-i\gamma L}+U_{15}U_{25} e^{-i\lambda L})+  \\
  &&  U_{12}U_{22}( U_{11}U_{21}e^{-i\delta L}+ U_{12}U_{22} + U_{13}U_{23} 
e^{-i\Delta L}+ \nonumber\\
  && U_{14}U_{24}e^{-i\gamma L}+U_{15}U_{25} e^{-i\lambda L})+ 
  U_{13}U_{23}( U_{11}U_{21}e^{-i\Delta L}+ U_{12}U_{22}e^{-i\Delta L}
 \nonumber \\
  && + U_{13}U_{23}+ U_{14}U_{24}e^{-i\gamma L}+U_{15}U_{25} e^{-i\lambda L}) 
    +U_{14}U_{24}((U_{11}U_{21}+U_{12}U_{22}
\nonumber\\
 && + U_{13}U_{23})e^{-i\gamma L}+U_{14}U_{24}+U_{15}U_{25}e^{-i\kappa L})
\nonumber \\
  &&+U_{15}U_{25}((U_{11}U_{21}+U_{12}U_{22}+ U_{13}U_{23}) e^{-i\lambda L}+
U_{14}U_{24}e^{-i\kappa L}+ U_{15}U_{25})] \nonumber
\eeq 
\vspace{3mm}

 From Eq(\ref{Pue-2})
\beq
\label{Pue-3}
 \mathcal{P}(\nu_\mu \rightarrow\nu_e) &=& U_{11}^2 U_{21}^2+
 U_{12}^2 U_{22}^2+ U_{13}^2 U_{23}^2+ U_{14}^2 U_{24}^2+U_{15}^2 U_{25}^2 + 
\nonumber \\
  &&  2U_{11} U_{21} U_{12} U_{22} cos\delta L + \\
  && 2(U_{11} U_{21} U_{13} U_{23}+ U_{12} U_{22} U_{13} U_{23})cos\Delta L+
\nonumber \\
  &&2U_{14}U_{24}(U_{11} U_{21}+ U_{12} U_{22}+ U_{13} U_{23})cos\gamma L+
\nonumber \\
  && 2U_{15}U_{25}(U_{11} U_{21}+ U_{12} U_{22}+ U_{13} U_{23})cos\lambda L+
 \nonumber \\
  && 2 U_{14}U_{24}U_{15}U_{25} cos\kappa L \nonumber \; .
\eeq 
\vspace{3mm}

  From the discussion below Eq(\ref{U1j}) , $\alpha \simeq \beta \simeq 9.2^o$,
with $sa=sb \simeq 0.16$ and $ca=cb\simeq 0.9871$,
 which are used to determine $U_{1j},U_{2j}$ in Eq(\ref{U1j}).

 In the figure below, the results of the two
sterile neutrinos on $\mathcal{P}(\nu_\mu \rightarrow \nu_e)$ using 
Eq(\ref{Pue-3}) and the parameters obtained from Ref\cite{collin16}
are shown for four experimental neutrino oscillation experiments. 

The figure also shows $\mathcal{P}(\nu_\mu \rightarrow \nu_e)$ with 
$\alpha=\beta= 0^o$, giving the results of a recent 3x3 S-mtrix 
calculation\cite{lsk14-2} to compare to the results with two sterile neutrinos.

\clearpage
\hspace{3cm}Using Eq(\ref{Pue-3}), one finds $\mathcal{P}(\nu_\mu \rightarrow 
\nu_e)$
\vspace{5.7cm}

\begin{figure}[ht]
\begin{center}
\epsfig{file=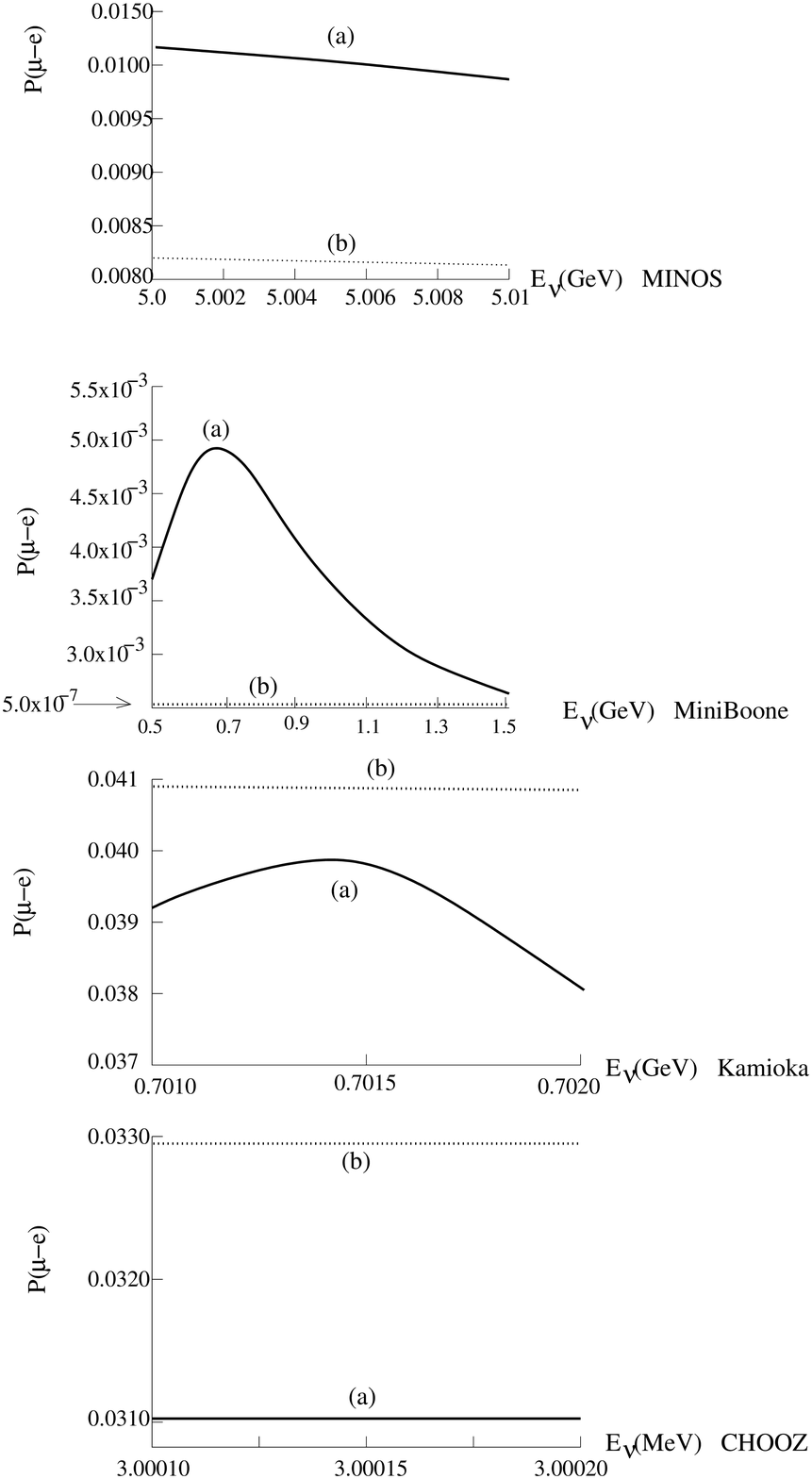,height=12cm,width=10cm}
\end{center}
\caption{$\mathcal{P}(\nu_\mu \rightarrow\nu_e)$ 
for MINOS(L=735 km), MiniBooNE(L=500m), JHF-Kamioka(L=295 km), and 
CHOOZ(L=1.03 km). (a) solid for $\alpha=\beta$=$9.2^o$;  
(b) dashed curve for $\alpha=\beta=\gamma$=$0^o$ giving the 3x3 result.}
\end{figure}
\clearpage
\section{$\mathcal{P}(\nu_\mu \rightarrow \nu_e)$ For One Sterile
Neutrino}

As mentioned above, in our previous articles on $\mathcal{P}(\nu_\mu 
\rightarrow \nu_e)$ with one sterile neutrino\cite{lsk14,lsk15} we used
the sterile neutrino-active neutrino mixing angle = $9.2^o$ 
from Refs\cite{kms11,kmms13,ggllz15}, while from the recent Ref\cite{collin16}
the sterile neutrino-active neutrino mixing angle $\alpha=\theta_{i4}
\simeq 9.2^o$, with $sa \simeq 0.16$ and $ca \simeq 0.9871$ for the first
sterile neutrino. Also, in Refs\cite{lsk14,lsk15} we used $\delta m_{4i}^2=
m_4^2-m_i^2 \simeq$ 0.9 (eV)$^2$, while from Ref\cite{collin16} $\delta m_{4i}^2
 \simeq 1.75 (eV)^2$, which also changes the estimate of $\mathcal{P}(\nu_\mu 
\rightarrow \nu_e)$ with one sterile neutrino. 
 
As discussed in Ref\cite{lsk14} the transition probability $\mathcal{P}(\nu_\mu 
 \rightarrow\nu_e)$, assuming $\delta_{CP}=0$ giving  $U^*_{ij}=U_{ij}$, is
\beq
\label{Pue}
 \mathcal{P}(\nu_\mu \rightarrow\nu_e) &=& U_{11}^2 U_{21}^2+
 U_{12}^2 U_{22}^2+ U_{13}^2 U_{23}^2+  \nonumber \\
  && U_{14}^2 U_{24}^2+ 2U_{11} U_{21} U_{12} U_{22} cos\delta L + \nonumber \\
  && 2(U_{11} U_{21} U_{13} U_{23}+ U_{12} U_{22} U_{13} U_{23})cos\Delta L+
\nonumber \\
  &&2U_{14}U_{24}(U_{11} U_{21}+U_{12} U_{22}+U_{13} U_{23})cos\gamma L \; ,
\eeq
with the parameters defined above.

Using  $c_{12}=.83,\;s_{12}=.56,\;s_{23}=c_{23}=.7071$, and $s_{13}=.15$,
(with $s_{ij}, c_{ij}=sin\theta_{ij},cos\theta_{ij}$),
\beq
\label{Uij}
   U_{11}&=& .822 c_\alpha \nonumber \\
   U_{12}&=&.554c_\alpha -.821 s_\alpha^2 \nonumber \\
   U_{13}&=&-.821s_\alpha^2c_\alpha-.554s_\alpha^2+.15c_\alpha  \nonumber \\
   U_{14}&=&.821s_\alpha c_\alpha^2+.554s_\alpha c_\alpha+.15s_\alpha  \\
   U_{21}&=& -.484c_\alpha  \nonumber \\
   U_{22}&=&.484s_\alpha^2 +.527 c_\alpha   \nonumber \\
   U_{23}&=&.699 c_\alpha-(-.484 s_\alpha c_\alpha+.527 s_\alpha)s_\alpha
\nonumber \\
   U_{24}&=&-.484 s_\alpha c_\alpha^2+.527s_\alpha c_\alpha+.699s_\alpha 
\nonumber \; ,
\eeq
with $\alpha$ the sterile-active neutrino mixing angle, $s_\alpha,c_\alpha$=
$sin(\alpha),cos(\alpha)$

In Figure 2 we compare $\mathcal{P}(\nu_\mu \rightarrow \nu_e)$ with one 
sterile neutrino using the sterile-active neutrino mixing angle of 
$9.2^0$\cite{collin16} ($sin(\alpha)\simeq 0.16$),
and  $\delta m_{4i}^2 \simeq 1.75 (eV)^2$ vs $\delta m_{4i}^2 \simeq 0.9 (eV)^2$
in Ref\cite{lsk15}; and $\mathcal{P}(\nu_\mu \rightarrow \nu_e)$ with no sterile
neutrino. Note the results are different fron those in Ref\cite{lsk15} because
of the mass differences.

\clearpage

  In Figure 2 the solid curves are estimates of $\mathcal{P}(\nu_\mu 
\rightarrow\nu_e)$ using the parameters from Ref\cite{collin16} for one
sterile neutrino, while the dashed curves are 
$\mathcal{P}(\nu_\mu \rightarrow\nu_e)$ with only active neutrinos.
\vspace{5cm}

\begin{figure}[ht]
\begin{center}
\epsfig{file=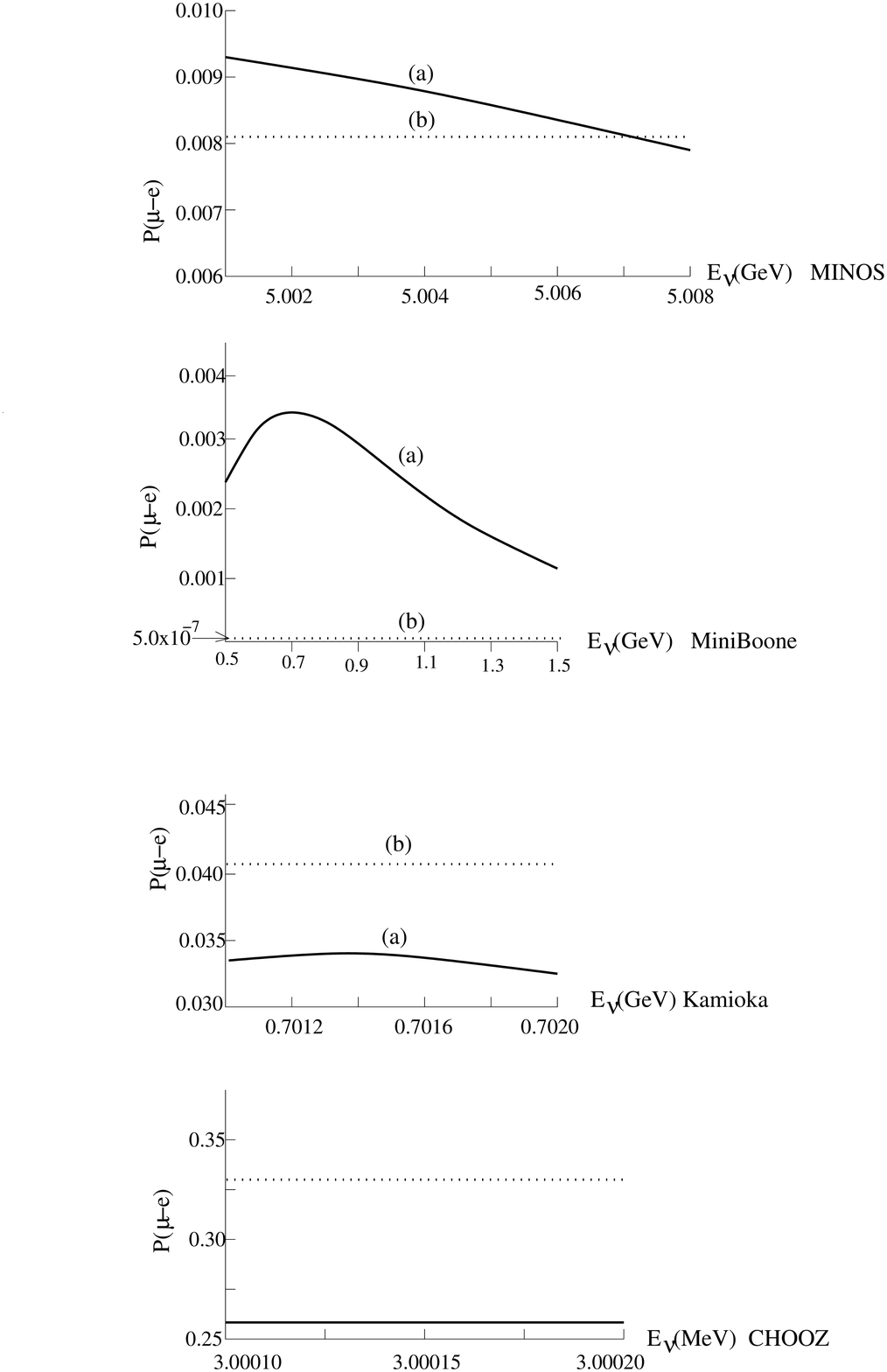,height=12cm,width=10cm}
\end{center}
\caption{$\mathcal{P}(\nu_\mu \rightarrow\nu_e)$ 
for MINOS(L=735 km), MiniBooNE(L=500m), JHF-Kamioka(L=295 km), and 
CHOOZ(L=1.03 km). (a) solid for $\alpha=9.2^o$;(b) dashed curve for 
$\alpha=0^o$, giving the 3x3 result.}
\end{figure}

\newpage
\section{Conclusions}

  From Figure 1 we note that even with the small mixing angles, $\alpha=\beta
=9.2^o$, obtained from the analyses given in Ref\cite{collin16} there is 
significant diference between our 5x5 and the earlier 3x3 prediction for 
$\mathcal{P}(\nu_\mu \rightarrow \nu_e)$, given by  $\alpha=\beta=0^o$. Also,
from Figure 2, for one sterile neutrino $\mathcal{P}(\nu_\mu \rightarrow \nu_e)$
differs significantly for $\alpha=9.2^o$, $\delta m_{4i}^2 \simeq 1.75 (eV)^2$
given in Ref\cite{collin16} compared to $\alpha=9.2^o$, $\delta m_{4i}^2 \simeq 
0.9 (eV)^2$ used in Ref\cite{lsk15}.

Therefore in future neutrino oscillation experiments the effect of two 
sterile neutrinos should be measured. Also the value of the sterile-active 
neutrino mixing angle might be more accurately determined in the near future.

\Large
{\bf Acknowledgements}
\vspace{3mm}

\normalsize
This work was carried out in part while LSK was a visitor at Los Alamos
National Laboratory, Group P25. The author thanks Dr. William Louis for 
several discussions and information concerning neutrino oscillation experiments
and the analysis to determine the mixing angles.

\end{document}